# High-pressure phase diagram of BaNi$_2$As$_2$: Unconventional charge density waves and structural phase transitions

Tom Lacmann,[1,*] Amir-Abbas Haghighirad,[1,†] Sofia-Michaela Souliou,[1] Michael Merz,[1,2] Gaston Garbarino,[3] Konstantin Glazyrin,[4] Rolf Heid,[1] and Matthieu Le Tacon[1,‡]

[1]*Institute for Quantum Materials and Technologies, Karlsruhe Institute of Technology, Kaiserstrasse 12, 76131 Karlsruhe, Germany*
[2]*Karlsruhe Nano Micro Facility, Karlsruhe Institute of Technology, Kaiserstrasse 12, 76131 Karlsruhe, Germany*
[3]*ESRF, The European Synchrotron, 71 Avenue des Martyrs, CS40220, 38043 Grenoble Cedex 9, France*
[4]*Deutsches Elektronen-Synchrotron DESY, Notkestrasse 85, 22607 Hamburg, Germany*

Structural phase transitions accompanied by incommensurate and commensurate charge density wave (CDW) modulations of unconventional nature have been reported in BaNi$_2$As$_2$, a nonmagnetic cousin of the parent compound of Fe-based superconductors, BaFe$_2$As$_2$. The strong dependence of the structural and CDW transitions of BaNi$_2$As$_2$ on isoelectronic substitutions alongside original dynamical lattice effects suggests strong tunability of the electronic phase of the system through structural effects. Here, we present a comprehensive synchrotron x-ray diffraction and first-principles calculation study of the evolution of the crystal structure and lattice instabilities of BaNi$_2$As$_2$ as a function of temperature and hydrostatic pressure (up to 12 GPa). We report a cascade of pressure-induced structural phase transitions and electronic instabilities up to ≈10 GPa, above which all CDW superstructures disappear. We reveal that the stable high-pressure phase consists of planar Ni zigzag chains, from which the surrounding As atoms have been pushed away. This yields a strong reduction of the interlayer As-As distance (along the original $c$ axis), akin to what is observed in the collapsed tetragonal structure of other pnictides, albeit here with a monoclinic structure. The discovery of polymorphs in the pressure-temperature phase diagram of BaNi$_2$As$_2$ emphasizes the importance of the relative Ni-Ni and Ni-As bond lengths in controlling the electronic ground state of this compound and increases our understanding of viable electronic phases under extreme conditions.

## I. INTRODUCTION

Superconductivity and charge density waves (CDWs) stand among the most commonly encountered instabilities of the metallic state and often coexist in the complex phase diagrams of quantum materials. Prominent examples include $\alpha$-U [1], high-temperature superconducting cuprates [2], transition-metal dichalcogenides [3], and the more recently discovered kagome superconductors [4]. Both electronic orders have also been evidenced in BaNi$_2$As$_2$, a weakly correlated metallic system which has, at room temperature, the same tetragonal $I4/mmm$ crystal structure as the parent compound of Fe-based superconductors, BaFe$_2$As$_2$ [5]. Upon cooling, rather than a magnetostructural transition, BaNi$_2$As$_2$ exhibits an original form of dynamical lattice nematicity [6,7] before undergoing a series of CDW instabilities and structural distortions [5,8–10] and ultimately entering a low-temperature superconducting phase below ∼0.6 K.

*Tom.Lacmann@kit.edu
†Amir-Abbas.Haghighirad@kit.edu
‡Matthieu.LeTacon@kit.edu

Incommensurate CDW (I-CDW) fluctuations have been associated with an enhanced elastoresistance signal in the $B_{1g}$ channel [11,12] and have been detected in thermal diffuse x-ray scattering already at room temperature [13] at the I-CDW wave vector $(\pm 0.28\,0\,0)_{\text{tet}}$ [note that throughout this paper, for simplicity, all reciprocal space indices are given in the tetragonal notation $(HKL)_{\text{tet}}$, but for better readability the subscript tet will generally be omitted]. A long-range I-CDW order develops fully only at ∼147 K [13] and triggers a minute orthorhombic distortion, bringing the system into an $Immm$ phase [7,10]. Below $T_{\text{tri}} \sim 137$ K (upon cooling) the system then undergoes a first-order transition to a triclinic ($P\bar{1}$) phase, while the I-CDW is replaced by a commensurate CDW (C-CDW) with a wave vector $(\pm 1/3\,0\,\mp 1/3)$ [5,9,10].

Various substitutions have been used and yield a rapid decrease of $T_{\text{tri}}$ alongside a sudden increase in the superconducting transition temperature $T_c$ to ∼3.5 K which occurs when the triclinic phase is completely suppressed [7,12,14]. Apart from the generic trends, details appear to be strongly dependent on the nature of the substitution. For instance ∼60% of strontium on the barium site is needed to suppress completely the triclinic phase [12], an effect obtained with only ∼7% of phosphorus substitution for arsenic [7,14,15] or ∼12% of cobalt on the nickel site [9,16]. All these substitutions are, in principle, isoelectronic and therefore do not change the charge carrier concentration in the system. On the other hand, these substitutions significantly affect the lattice parameters [7,10,14,17] in a nontrivial manner, which suggests in turn, akin to Fe-based superconductors [18], that pressure might be a valuable tuning parameter of the electronic phase of $BaNi_2As_2$. To the best of our knowledge, high-pressure investigations on this system have been limited to only resistivity measurements of pristine $BaNi_2As_2$ [19] in a limited pressure range (up to 2.74 GPa) where only a modest dependence of both structural and superconducting transition temperatures was found.

Here, we use high-resolution single-crystal x-ray diffraction (XRD) to investigate the pressure and temperature dependence of the various structural phases of $BaNi_2As_2$ over a broader pressure range and to construct a temperature-pressure phase diagram of this compound extending up to 12 GPa and down to 30 K. We carried out systematic structural refinement and discovered a series of pressure-induced structural phase transitions, as well as a set of superstructures associated with different CDW instabilities (incommensurate and commensurate). These instabilities show a highly unusual pressure dependence but are well described by our first-principles density-functional perturbation theory (DFPT) calculations, which also emphasize the absence of Fermi surface nesting associated with these instabilities and only weak electron-phonon coupling enhancement of the phonon line shapes (if any). This is in sharp contrast to the phenomenology encountered in prototypical CDW systems and indicates the unconventional nature of all these CDW instabilities. Above ∼10 GPa, all superstructure peaks disappear, and our study reveals that the stable high-pressure phase is monoclinic $C2/m$ and consists of planar Ni zigzag chains. Akin to what is observed in the collapsed tetragonal (cT) phase of other pnictide compounds, an overall decrease in the out-of-plane $c$ axis parameter is observed as the As-As distance between the Ni planes is strongly reduced [20,21]. On the other hand, in contrast to known cT phases, the "thickness" of the NiAs layers increases in the high-pressure phase compared to that of the tetragonal $I4/mmm$ structure at lower pressures, as the As atoms are pushed away from the Ni planes. This peculiarity emphasizes the importance of the hybridization between the As $4p$ and Ni $3d$ orbitals, through the Ni-As and As-As distances, in controlling the electronic phase of $BaNi_2As_2$.

## II. EXPERIMENTAL DETAILS

### A. Single-crystal growth and characterization

Single crystals of $BaNi_2As_2$ were grown using a self-flux method. NiAs precursor was synthesized by mixing the pure elements Ni (powder, Alfa Aesar 99.999%) and As (lumps, Alfa Aesar 99.9999%) that were ground and sealed in a fused silica tube and annealed for 20 h at 730 °C. All sample handlings were performed in an argon glove box ($O_2$ content < 0.1 ppm). For the growth of $BaNi_2As_2$, a ratio of Ba:NiAs = 1:4 was placed in an alumina tube, which was sealed in an evacuated quartz ampule ($10^{-5}$ mbar). The mixtures were heated to 700 °C for 10 h and then slowly heated to a temperature of 1090 °C, soaked for 5 h, and subsequently cooled to 995 °C at a rate of 0.25 °C/h. At 995 °C, the furnace was canted to remove the excess flux, followed by furnace cooling. Platelike single crystals with typical sizes of $3 \times 2 \times 0.5$ mm$^3$ were easily removed from the ingot. The crystals are shiny brass yellow with a metallic luster. All samples were selected from the same batch and cleaved with a scalpel just before the x-ray scattering experiments.

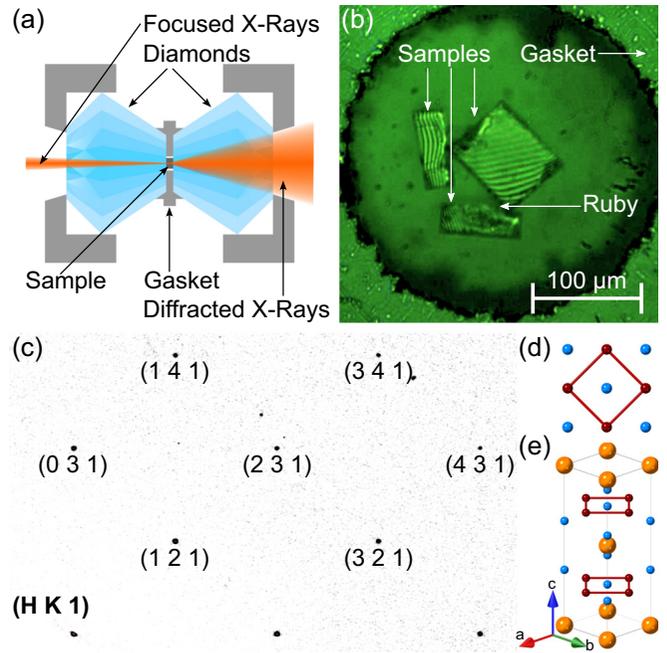

FIG. 1. Experiment overview. (a) Schematics of a diamond anvil cell and scattering geometry used for the HP-LT XRD experiments. (b) Enlarged view of the inner assembly of the DAC containing $BaNi_2As_2$ single crystals, a ruby single crystal (as a pressure marker), and the gasket. The samples are positioned with the crystallographic $c$ direction parallel to the beam at ≈0° and the diamond surface normal. The edges of the samples correspond to $a$ or $b$ lattice directions. (c) Observed x-ray diffraction patterns for a $BaNi_2As_2$ single crystal at 196 K and 0.29 GPa shown for a $(HK1)$ plane. (d) Top view on the Ni-As plane and (e) the $BaNi_2As_2$ unit cell as refined with the data shown in (c). The barium atoms are shown in orange, the nickel atoms are in red, and the arsenide atoms are in blue.

### B. High-pressure x-ray diffraction

High-pressure, low-temperature experiments were performed at the European Synchrotron Radiation Facility (ESRF, beamline ID15B) and Positron-Elektron-Tandem-Ring-Anlage III (PETRA III, DESY, beamline P02.2) in a membrane-type diamond anvil cell (DAC) using the ruby fluorescence method for the pressure calibration [22]. For the experiments at the ESRF an ESRF-made Le Toullec-type CuBe-alloy DAC with diamonds with cullet diameters of 500 µm was used. A stainless-steel gasket was indented down to a thickness of ≈100 µm. $BaNi_2As_2$ singles crystals and one ruby were placed inside the gasket hole, and helium was used as the pressure-transmitting medium. A sketch and a photograph of the DAC, samples, ruby, and gasket hole after gas loading are shown in Figs. 1(a) and 1(b). For x-ray diffraction a monochromatic beam with an energy of 30.17 keV (≈0.411 Å) was focused down to $4 \times 4$ µm$^2$, and the diffracted beam was detected with a Mar research MAR555 flat panel detector. For each dataset images within an angular range of ±32° were recorded. Each image was integrated for 0.5 s and over a 0.5° rotation. The detector position and distance (259.9 mm) were calibrated with silicon powder and an enstatite single-crystal standard using the DIOPTAS [23] and CRYSALISPRO softwares [24]. For the measurements at PETRA III, a DESY-made hardened steel symmetric DAC with a diamond cullet diameter of 400 μm was used. A rhenium gasket was indented down to a thickness of ≈80 μm. The cells were loaded with neon as the pressure-transmitting medium. X-ray diffraction experiments were performed using a monochromatic x-ray beam with an energy of 42.71 keV (≈0.2903 Å) focused down to 3×8 μm$^2$. The diffracted beam was detected with a Perkin Elmer XRD 1621 detector. The detector position and distance (401.176 mm) were calibrated with CeO$_2$ and an enstatite single-crystal standard using the DIOPTAS [23] and CRYSALISPRO [24] softwares. At each pressure and temperature, images were recorded during a continuous rotation from −25° to 30°. Each image was integrated for 0.5 s over a 0.5° rotation.

TABLE I. Overview of the unit cell; space group; lattice parameters $a$, $b$, and $c$; angles $\alpha$, $\beta$, and $\gamma$; and $R$ factors for all six observed phases. For each dataset the corresponding pressure and temperature are stated. An extended version of this table can be found in the SM [25].

|  | Tetragonal | Orthorhombic | Triclinic | Monoclinic I | Monoclinic IIa | Monoclinic IIb |
|---|---|---|---|---|---|---|
| Pressure (GPa) | 0.21 | 0.29 | 0.14 | 4.6 | 8.64 | 11.39 |
| Temperature (K) | 192 | 146 | 128 | 94 | 145 | 94 |
| Space Group | $I4/mmm$ | $Immm$ | $P\bar{1}$ | $C2/c$ | $C2/m$ | $C2/m$ |
| $a$ (Å) | 4.1275(1) | 4.1227(14) | 4.1380(8) | 11.697(13) | 11.434(19) | 10.882(4) |
| $b$ (Å) | 4.1275(1) | 4.1230(12) | 4.1625(8) | 4.1210(15) | 4.0901(3) | 4.0874(1) |
| $c$ (Å) | 11.606(6) | 11.614(7) | 6.385(4) | 8.235(2) | 4.093(23) | 4.0961(45) |
| $V$ (Å$^3$) | 197.74(1) | 197.43(1) | 98.64(7) | 374.6(3) | 178.52(3) | 168.3(1) |
| $\alpha$ (°) | 90 | 90 | 108.47(3) | 90 | 90 | 90 |
| $\beta$ (°) | 90 | 90 | 107.695(3) | 109.29(7) | 111.17(12) | 112.48(26) |
| $\gamma$ (°) | 90 | 90 | 90.78(16) | 90 | 90 | 90 |
| $R_{int}$ | 0.0281 | 0.0180 | 0.0204 | 0.0210 | 0.0104 | 0.0254 |
| $R_1$ | 0.0170 | 0.0243 | 0.0279 | 0.0299 | 0.0258 | 0.0302 |

We will focus here on isotherms that were measured mainly by cooling from room temperature to the desired temperatures, at which we compressed the crystals to the target pressure. In the following, data from three different samples and three different DAC loadings are shown. Sample 1 was measured at the ESRF, while the samples 2 and 3 were measured at PETRA III, DESY. There was no noticeable difference between different samples or different sample positions. More details on the samples, the exact $p$-$T$ measurement path for each sample, and additional data obtained along five isobars (at ∼2.2, 4, 7.6, 10, and 12 GPa) can be found in the Supplemental Material (SM) [25]. As an example, the $(HK1)$ plane from the measurement at 0.29 GPa and 194 K is shown in Fig. 1(c), and the structure determined from a structural refinement of the corresponding dataset (space group $I4/mmm$) is shown in Figs. 1(d) and 1(e).

### C. Analysis of the crystal structure

CRYSALISPRO [24] was used for cell refinement, data reduction, and the analysis of the diffraction precession images for all datasets. SHELLXS97 [26] and SHELLXL97 2014/7 [27] as well as JANA2006 [28] were used to solve the crystal structure and refinements. Crystal data and structural refinement details are summarized in Table I and in the SM [25]. The x-ray crystallographic coordinates of the crystal structures we found have been deposited at the Cambridge Crystallographic Data Centre (CCDC) under deposition numbers 2289119–2289121. Atomic coordinates and site labels were standardized using the VESTA [29] crystal structure visualization software.

### III. COMPUTATIONAL DETAILS

Density-functional investigations of lattice dynamics properties for the different structural phases of BaNi$_2$As$_2$ were performed in the framework of the mixed-basis pseudopotential method [30,31]. This approach employs an efficient description of more localized components of the valence states by using a basis set combining plane waves and local functions at atomic sites. The electron-ion interaction is described by norm-conserving pseudopotentials, which were constructed following the descriptions of Hamann *et al.*

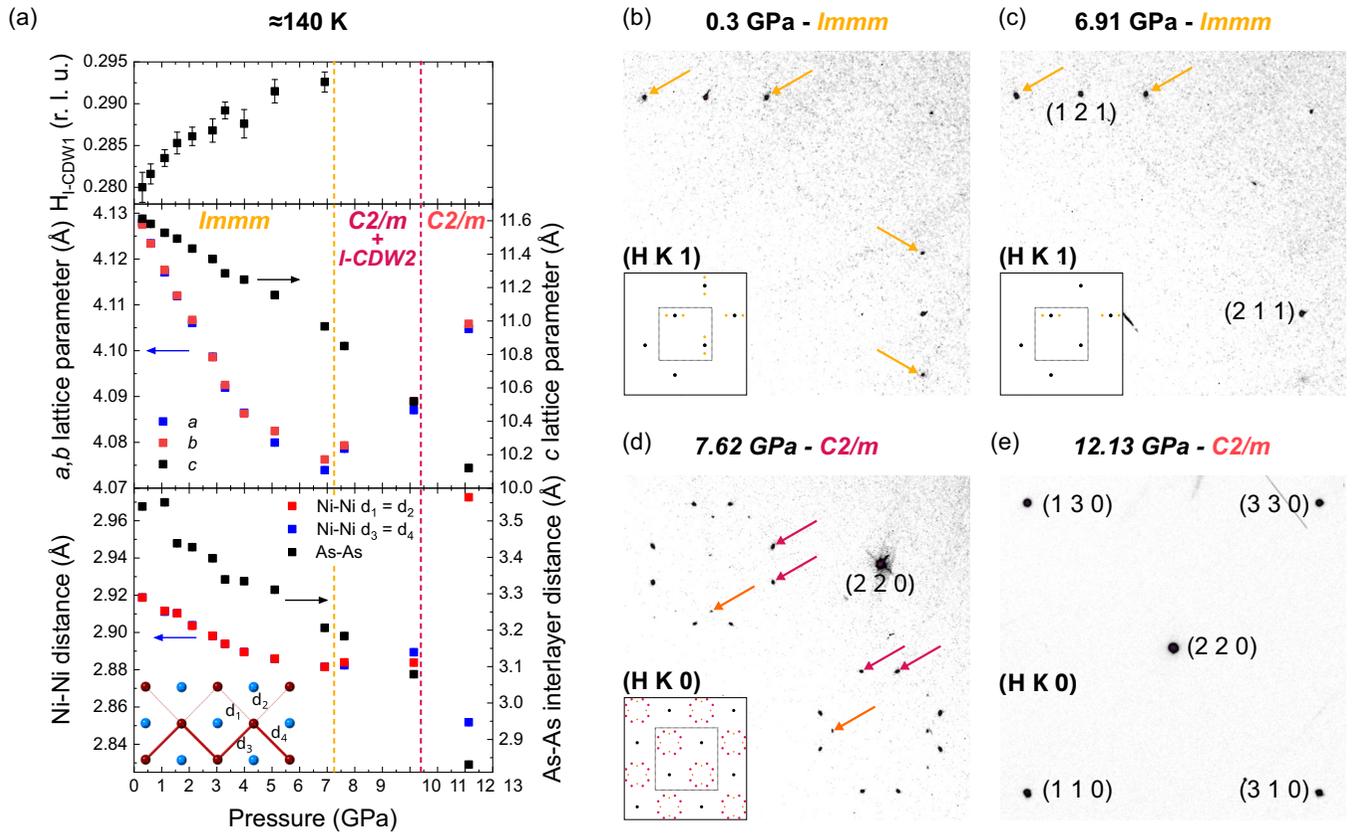

FIG. 2. I-CDW evolution with pressure: 140 K isotherm. (a) Pressure dependence of the I-CDW1 wave vector, lattice parameters, and Ni-Ni intralayer and As-As interlayer distance at 140 K. The lattice parameters are extracted using the orthorhombic *Immm* unit cell. A sketch of the Ni-As plane indicates the labeling of the different Ni-Ni distances. The orthorhombic *Immm* to monoclinic $C2/m$ structural transition as it is seen by the change in the $a$ and $b$ lattice parameters is indicated. Overview of the reciprocal space and observed CDWs at 140 K and (b) 0.3, (c) 6.91, (d) 7.62, and (e) 12.13 GPa in the ($HK1$) [(b) and (c)] and ($HK0$) [(d) and (e)] planes. The superstructure peaks from I-CDW1 (light orange), I-CDW1′ (dark orange), and I-CDW2 (magenta) are indicated by arrows. For each pressure a sketch of the reciprocal space showing the Bragg and superstructure peaks is included. The dotted square indicates the shown experimental data. All data shown are measured on sample 1 except for the data at 12.13 GPa in (a) and (e), which are measured on sample 3 and at 145 K.

[32,33] for Ba and Vanderbilt [34] for Ni and As. Semicore states Ba 5$p$, Ni 3$s$, and Ni 3$p$ were included in the valence space. The exchange-correlation functional was represented by the general gradient approximation in the Perdew-Burke-Ernzerhof form [35]. The mixed-basis set consisted of plane waves with a cutoff for the kinetic energy of 22 Ry and local functions of $p$ and $d$ type for Ba and $s$, $p$, and $d$ type for Ni. Lattice dynamics properties were calculated within the linear response of DFPT as implemented in the mixed-basis method [36]. Brillouin zone integration was performed by sampling a $16 \times 16 \times 8$ $k$-point mesh in conjunction with a Gaussian broadening of 0.1 eV. To locate positions of phonon anomalies in momentum space, scans of the phonon dispersions on two-dimensional high-symmetry planes were performed as follows. Dynamical matrices were calculated within DFPT on an $8 \times 8$ mesh and interpolated on a much denser $120 \times 120$ mesh using a standard Fourier interpolation technique. Diagonalizing the dynamical matrices provided phonon frequencies.

## IV. RESULTS AND DISCUSSION

In this section we present evidence for the existence of different HP phases in BaNi$_2$As$_2$ which can be best seen following two isotherms at 140 K (above the triclinic transition at ambient pressure) and 94 K (below the triclinic transition at ambient pressure). Up to about 10 GPa, each of these HP phases is accompanied by a different type of CDW modulation. Above this pressure, CDW superstructures disappear.

### A. Pressure dependence of the I-CDW: 140 K isotherm

As previously discussed [7,10,13], the formation of the long-range I-CDW at ambient pressure (hereafter referred to as I-CDW1) is accompanied by a fourfold symmetry-breaking transition. This is best seen as a difference between the thermal expansion along the (100) and (110) directions [7,10] and indicates a small but measurable orthorhombic distortion below $\sim$146 K. Consequently, we can index the Bragg reflections obtained at 140 K and close to ambient pressure in a slightly distorted orthorhombic cell with space group *Immm*. The corresponding structural parameters are detailed in Table I. In agreement with previous reports [10,13,17], I-CDW1 satellites are observed around Bragg reflections at ($\pm 0.28$ 0 0) or (0 $\pm 0.28$ 0), depending on the reflection [see Figs. 2(b) and 2(c)].

The effect of pressure on the unit cell is reported in Fig. 2(a), where we show that the *a*, *b*, and *c* lattice parameters decrease smoothly with increasing pressure up to 7 GPa. As the orthorhombicity increases upon pressurization, half of the I-CDW1 superstructure peaks disappear [see Fig. 2(c)]. The latter can be interpreted as a consequence of detwinning that could originate either from weak nonhydrostaticity in the pressure cell or from the anisotropic response of BaNi$_2$As$_2$ to strain. In parallel, we observe an increase in the incommensurability of I-CDW1 from 0.28 at ambient pressure to 0.293 at 7 GPa [Fig. 2(a)]. Above ∼7 GPa the I-CDW1 satellites disappear, and a new set of eight incommensurate satellites appears close to wave vectors (±0.358 ±0.10 0) and (±0.10 ±0.358 0) around, e.g., the (220) Bragg peaks [Fig. 2(d)], forming a new I-CDW, labeled I-CDW2 hereafter. Furthermore, I-CDW2 shows a strong temperature dependence of the wave vector and begins at a higher temperature of ≈168 K. Details can be found with the evaluation of the isobars around 2.2, 4, and 7.6 GPa in the SM [25]. Note that although the original I-CDW1 peaks are lost above 7 GPa, some faint peaks with similar wave vectors can still be seen up to 10 GPa, albeit now centered around forbidden Bragg reflections such as the (120) reflection (denoted as I-CDW1′). Additionally, the pressure dependence of *a* and *b* lattice parameters displays a sudden upturn, indicating that a structural phase transition takes place between 7 and 7.6 GPa. The smooth evolution of the lattice parameters and absence of a coexistence regime suggest that this is a second-order phase transition.

Our structural refinement in this region shows that the crystal structure is monoclinic and can be described within the space group *C*2/*m*. This structural phase transition involves small atomic displacements that break the translational symmetry of the lattice (or, equivalently, domain-related distortions as the minute difference between cell parameters *a* and *b* in the *Immm* structure increases on approaching the monoclinic phase) and a shear displacement of the Ni layers against each other. This amounts to a loss of symmetries of both the As and Ni sites as additional degrees of freedom are introduced in Wyckoff position 4*i* by breaking the correlation between the *x* and *z* components at this position. The 4*i* site symmetry then changes from *mm*2 in the *Immm* phase to *m* in the *C*2/*m* phase. In this phase, instead of four equivalently long Ni-Ni bonds, regular Ni zigzag chains with two long and two short bond distances form [bottom panel of Fig. 2(a)].

Above ∼10.2 GPa, all CDW satellites completely disappear [see Fig. 2(e)]. Although the symmetry remains the same as the I-CDW2 disappears, a closer look at the crystal structure reveals important internal changes [Fig. 2(a)] both in and out of the NiAs planes. In plane, the Ni-Ni bond length disproportionation strongly increases (reaching 4.2% at 12 GPa), indicating an increased separation of the zigzag chains. In the perpendicular direction (*c* axis in the *I*4/*mmm* setting), the As-As distance between NiAs layers $d_{\text{As-As}}$ decreases continuously with increasing pressure, reaching a value of 3.079(9) Å at 10 GPa [Fig. 2(a)]. Above 10 GPa, $d_{\text{As-As}}$ abruptly reduces to 2.831(7) Å, which is reminiscent of the first-order transition to cT phases in other iron [37] and cobalt [38] pnictide families. The jump in the internal parameters indicates that the transition to the monoclinic IIb phase is first order. In this phase, the As-As distance within the NiAs layers increases (or, equivalently, the As-Ni-As angle decreases), showing that the NiAs layers become thicker in the high-pressure phase. This can be evidenced only by looking carefully at the bond distances since, overall, the unit cell size perpendicular to the Ni planes decreases. The isobar measurements at ∼12 GPa indicate the absence of transition upon cooling at this pressure between 200 and 50 K [25].

### B. Pressure dependence of the C-CDW: 94 K isotherm

Next, we look at the impact of pressure on the triclinic phase of BaNi$_2$As$_2$ where the C-CDW is seen at ambient pressure down to the lowest temperatures [5,7]. Previous studies indicated that in this phase the four Ni-Ni bond distances become nonequivalent, forming in-plane Ni-Ni dimers [8,10,39], as can be derived from the different Ni-Ni bond distances in Fig. 3(a).

Before discussing the evolution of the crystal structure, let us first focus on the pressure dependence of the C-CDW superstructure (we recall here that for simplicity, the corresponding reflections are indexed in the tetragonal setting). In the triclinic phase, the characteristic set of C-CDW satellites with wave vectors (±1/3 0 ∓1/3) is still clearly visible in the (*H*2*L*) plane at 2.11 GPa [Fig. 3(b)]. A new set of C-CDW (C-CDW2) superstructure peaks is observed at 3.75 GPa around wave vectors (±1/2 0 ∓1/2). Note that at this pressure, weak signatures of the C-CDW1 satellites can still be seen, indicating a narrow coexistence region of the two orders. The C-CDW1 satellites are completely suppressed with increasing pressure, while the C-CDW2 peaks remain visible up to 9.5 GPa. Above 10 GPa, like for the 140 K isotherm, no superstructure reflections could be observed [see Fig. 3(e)].

From a structural point of view, as previously discussed, the high-pressure phase above 10 GPa is best described as a "collapsed" monoclinic *C*2/*m* structure, with a reduced As-As distance for the As ions connecting the Ni-As layers. In contrast to the situation at higher temperatures, however, evaluating the structure by including the C-CDW2 modulation in the refinement yields poor results when describing it in the *C*2/*m* space group. On the other hand, it is quite clear as well that the triclinic $P\bar{1}$ phase is suppressed alongside the C-CDW1 phase above 2.5 GPa. The best structural solution for this intermediate pressure phase (i.e., between 3.75 and 9.5 GPa) is obtained by including the C-CDW2 superstructure reflections explicitly to solve the structure in the monoclinic *C*2/*c* space group (this corresponds in particular to a doubling of the unit cell in the plane, in which three inequivalent Ni-Ni bonds are found). The transition between the two monoclinic phases with *C*2/*m* and *C*2/*c* space groups occurs around 10 GPa at 94 K, where the monoclinic *β* angle and the ratio of the *a* and *b* lattice parameters [Fig. 3(a)] exhibit clear discontinuities. All these transitions are first order in nature, and in contrast to the situation at higher temperatures where the symmetry of the unit cell was lowered with increasing pressure, symmetries are restored under pressurization at low temperature.

### C. Pressure-temperature phase diagram and discussion

We illustrate the results of our analysis of the crystal structures and superstructures of BaNi$_2$As$_2$ for each of the >100

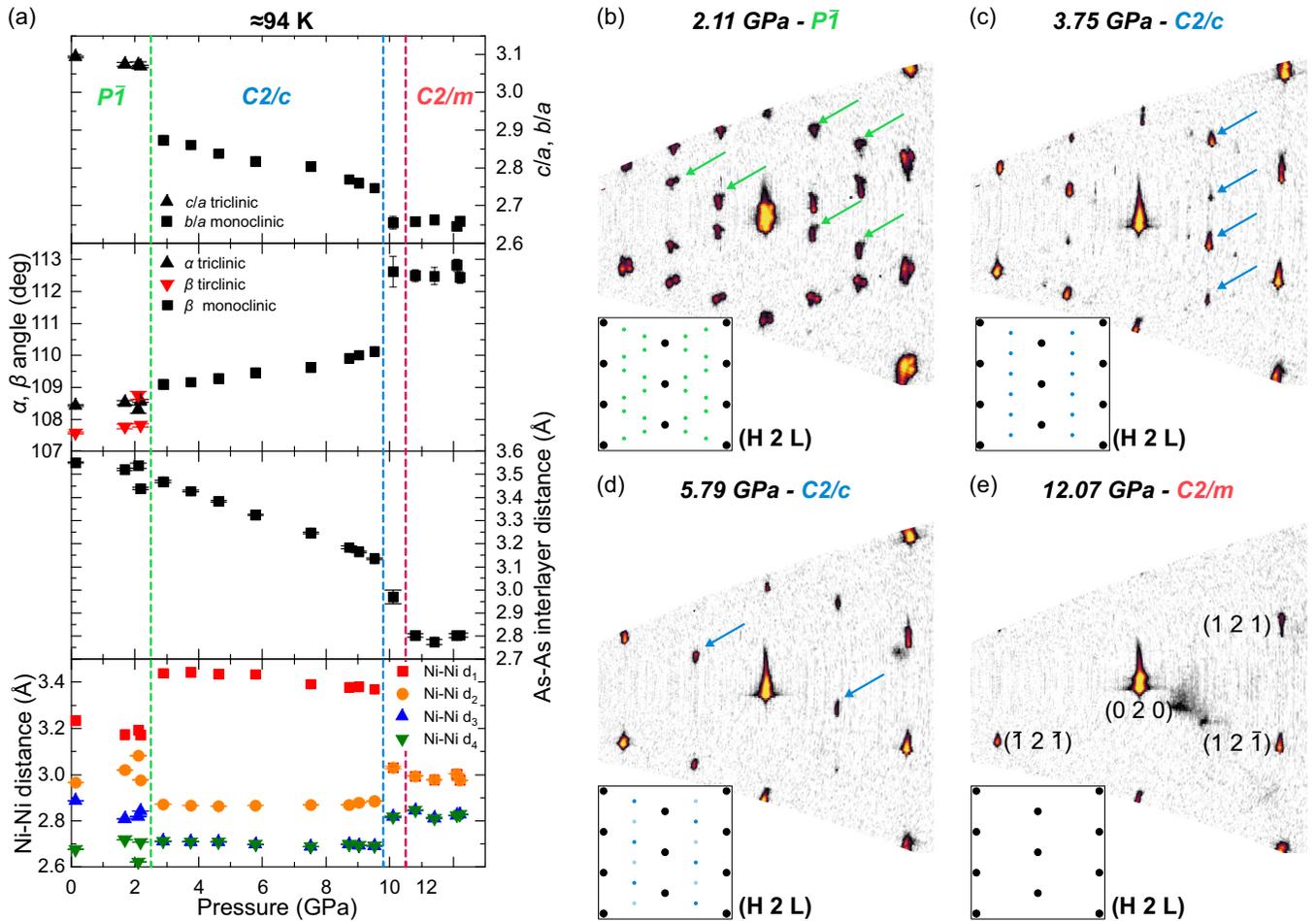

FIG. 3. Isotherm at 94 K. (a) Pressure dependence of the $c/a$ or $b/a$ ratio, triclinic or monoclinic angles $\alpha$ and $\beta$, and As-As interlayer and Ni-Ni intralayer distances. The $c/a$ and $b/a$ ratios and angles $\alpha$ and $\beta$ indicate similar quantities in the different structures. The transitions from $P\bar{1}$ to $C2/c$, $C2/m$ with I-CDW2, and $C2/m$ without I-CDW2 are indicated by dashed vertical lines. An overview of the reciprocal space around the (020) Bragg peak for (b) 2.11, (c) 3.75, (d) 5.79, and (e) 12.07 GPa at 94 K. The superstructure peaks from C-CDW1 (light green) and C-CDW2 (light blue) are indicated by arrows. For each pressure a sketch of the reciprocal space showing the Bragg and superstructure peaks is included. The data shown in (b)–(e) were measured on sample 3. Measurements on samples 1 and 2 are also included in (a). For details on which sample was measured at which $p$-$T$ point, see the SM [25].

pressure-temperature points measured in a detailed phase diagram in Fig. 4. Crystallographic parameters for each structure are given in Table I and in the SM [25]. It is worth mentioning that for all the phases the crystallographic refinements are performed without including the CDW reflections, except for the case of C-CDW2 in the monoclinic I phase.

The first important observation is that the phase diagram shows a qualitatively different pressure dependence for the high (orthorhombic and I-CDW) and low (triclinic and C-CDW) temperature phases. While the low-temperature triclinic phase is lost already between 2 and 3 GPa, the orthorhombic phase around 140 K survives up to ∼7 GPa. We have seen that this is accompanied by a continuous change in the incommensurability of the I-CDW1 with increasing pressure, whose onset temperature, nonetheless, does not seem to strongly vary with pressure. This is also the case for the first-order transition temperature to the triclinic/C-CDW1 phase below ∼3 GPa, in agreement with an earlier transport study [19]. This independence of the CDW formation temperatures over large pressure ranges contrasts significantly with the previously reported effects of chemical substitutions. This is particularly true in the case of the C-CDW and triclinic phases, which are gradually suppressed through substitution, e.g., by phosphorus, cobalt, or strontium [6,7,17]. This can be best understood by looking at the effect of these substitutions on the structure. Substitutions tend to have opposite effects in and out of plane, in contrast to hydrostatic pressure, which compresses all lattice parameters. The main effect of P or Co substitutions which efficiently suppress the triclinic and C-CDW1 phase is a contraction of the $ab$ plane [7,10,14,16]. On the contrary, Sr mostly induces a compression of the $c$ axis [12]. Interestingly, Sr substitution also induces a commensurate CDW with a doubling of the unit cell [17], which bears similarities to the one reported here, even though it has not been associated with a structural phase transition to a $C2/c$ phase. Its observation in the Sr-substituted samples—and not in the P- and Co-substituted samples—reemphasizes the importance of the out-of-plane structural details in controlling the electronic phases of the pnictides, akin to their Fe-based counterpart [40,41].

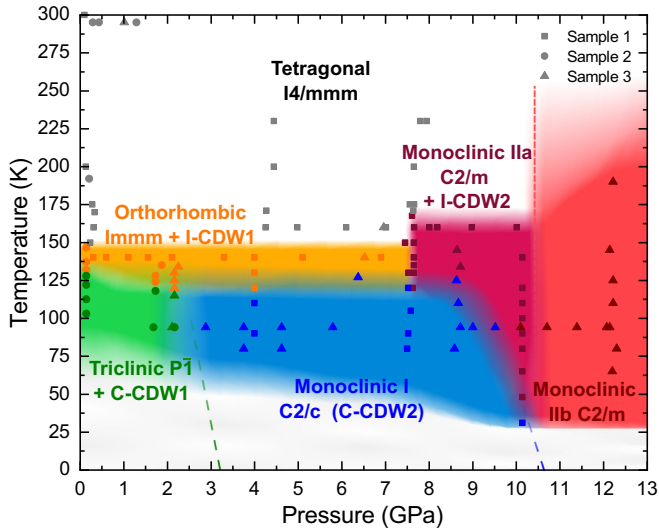

FIG. 4. Pressure-temperature phase diagram of BaNi$_2$As$_2$. The measured and refined pressure-temperature points of each of the three different samples are annotated by different symbols. The symbol color corresponds to the refined structure of each measurement. The phase spaces of the different structures and their associated CDWs are indicated by the background color. Except for the monoclinic I structure, for all phases the superstructures were not included in the refinement.

We note that the stability of the I-CDW1 up to 7 GPa contrasts with the observations in the vast majority of known CDW materials. For instance, in rare earth tritellurides $R$Te$_3$ [42,43], dichalcogenides such as 2$H$-NbSe$_2$ [44] and TiSe$_2$ [45], $\alpha$-U [46], and kagome superconductors [47], to name a few, the CDW ordering temperature is strongly dependent on pressure and most often decreases rapidly with increasing pressure, generally resulting in a complete suppression of the CDW after a few gigapascals. There are, of course, notable exceptions to this, such as VSe$_2$ [48] and SmNiC$_2$ [49], but in that case the CDW formation temperature rapidly increases with pressure. In this respect and to the best of our knowledge, the resilience of I-CDW1 in BaNi$_2$As$_2$ versus pressure is particularly remarkable. It might be related to the nematic liquid phenomenology evidenced at higher temperature in this compound [6] as a consequence of strong fluctuations between degenerate nematic configurations which is expected to be weakly affected by strain. Interestingly, the dramatic changes in the superstructure reflections, associated with CDWs in this work in analogy with the ambient pressure case, are concomitant with pressure-induced structural phase transitions. The formation of Ni zigzag chains yielding a monoclinic $C2/m$ structure above 7 GPa at high temperatures and a $C2/c$ structure above 3 GPa at low temperatures is associated with a remarkable change in the superstructure pattern, indicating a profound interdependence of the CDW instabilities and the underlying lattice structure. To gain further insights, we turn now to first-principles calculations, which are particularly favorable owing to the weakly correlated nature of the material.

The structures reported in this study have not been anticipated by previous theoretical investigations [39,50] because it is generally challenging to determine *a priori* the symmetry of the most stable structural configuration of a given compound. It is, nonetheless, possible to assess the stability of the experimentally determined crystal structures by looking at their lattice dynamics. Using the experimental lattice parameters and relaxed atomic positions to obtain force-free configurations prior to the phonon calculations, we previously showed that the dispersion of the phonons of the $I/4mmm$ tetragonal structure was unstable due to the presence of the softening of a low-lying optical phonon (dispersing from the Raman-active $E_g$ mode at the zone center) along the reciprocal ($H$00) direction and at a wave vector close to that of the experimental I-CDW1. We have extended this approach to the pressurized unit cells. In the color plots in Fig. 5, we map the lowest phonon frequencies (full dispersions are shown in the SM [25]) across planes of the reciprocal space. As unstable modes are characterized by imaginary frequencies, the negative modulus of the frequency was used so that the dominant instabilities show up as minima of the softest phonon frequency in Fig 5. In agreement with previous work, the calculation performed on the weakly distorted orthorhombic $Immm$ structures determined at 0.3 and 5.1 GPa indicates that the leading phonon instability occurs at (0.25, 0, 0) and (0, 0.25, 0), close to the I-CDW1 wave vector. Interestingly, at both pressures, we can already observe a weak softening of the same phonon branch close to the wave vector at eight locations in the ($HK$0) plane, including, e.g., (0.38 0.1 0) and (0.1 0.38 0), which are very close to that at which the I-CDW2 satellites (Fig. 2) have been observed. This becomes the leading instability at 10.14 GPa calculated within the monoclinic $C2/m$ phase [Fig. 5(c)], while upon further compression the phonon anomalies are suppressed and this phase is stabilized, as evidenced by the disappearance of negative phonon energies in Fig. 5(d).

Next, we discuss the instabilities of the low-temperature phases. In the DFPT calculation the experimental triclinic $P\bar{1}$ structure [shown for 1.69 GPa in Figs. 5(e) and 5(f)] is also found to be unstable due to the same low-lying phonon branch, but the leading instability is in this case found in the ($H0L$) plane. It is rather spread in the reciprocal space but centered around the (1/3 0 1/3) and (2/3 0 2/3) wave vectors, at which C-CDW1 satellites are seen experimentally [Fig. 3(b)]. Similarly, the leading instability of the $C2/c$ monoclinic structure at 5.79 GPa shown in Figs. 5(g) and 5(h) occurs at the commensurate wave vector (1/2 0 1/2) of the C-CDW2 phase [Figs. 3(b) and 3(c)].

In all these cases, similar to investigations at ambient pressure [13], no Fermi surface nesting is found at the I-CDW1, I-CDW2, C-CDW1, and C-CDW2 wave vectors (details are presented in the SM [25]). In general, we do not observe any anomaly in the phonon linewidth associated with the momentum structure of the electron-phonon coupling vertex that correlates with the structure of these CDWs, indicating the unconventional natures of these CDWs. The only noticeable exception is the I-CDW2 case, for which a weak enhancement of the linewidth of the unstable phonon is seen, but the calculated electron-phonon coupling remains very modest. It typically amounts to ∼0.15 meV, which is almost an order of magnitude weaker than in that of prototypical CDW systems such as dichalcogenides [51,52]. To sum up, whether the lattice structure of BaNi$_2$As$_2$ is stable with respect to CDWs seems to be completely controlled by the local environment of Ni and thus by the orbital polarization of the bands crossing the Fermi level, which originates primarily from the Ni states [39,50]. On the one hand, it is clear that the deformation of the Ni square lattice into a zigzag structure with a bond length disproportionation can occur only alongside a spectral weight transfer between the in- and out-of-plane $t_{2g}$ orbitals of Ni [10]. On the other hand, the main player causing the disappearance of CDWs above 10 GPa is the As atoms surrounding the planar Ni zigzag chains, which are pushed away as the interlayer As-As distance is strongly reduced. Our first-principles calculation indicates that the electron-phonon interaction is extremely sensitive to the subtle details of the hybridization between the As $4p$ and Ni $3d$ orbitals, primarily controlled by the Ni-As and As-As distances. Despite the weak spectral weight of As states at the Fermi level, they seem to play a key role in controlling the electronic phase of BaNi$_2$As$_2$.

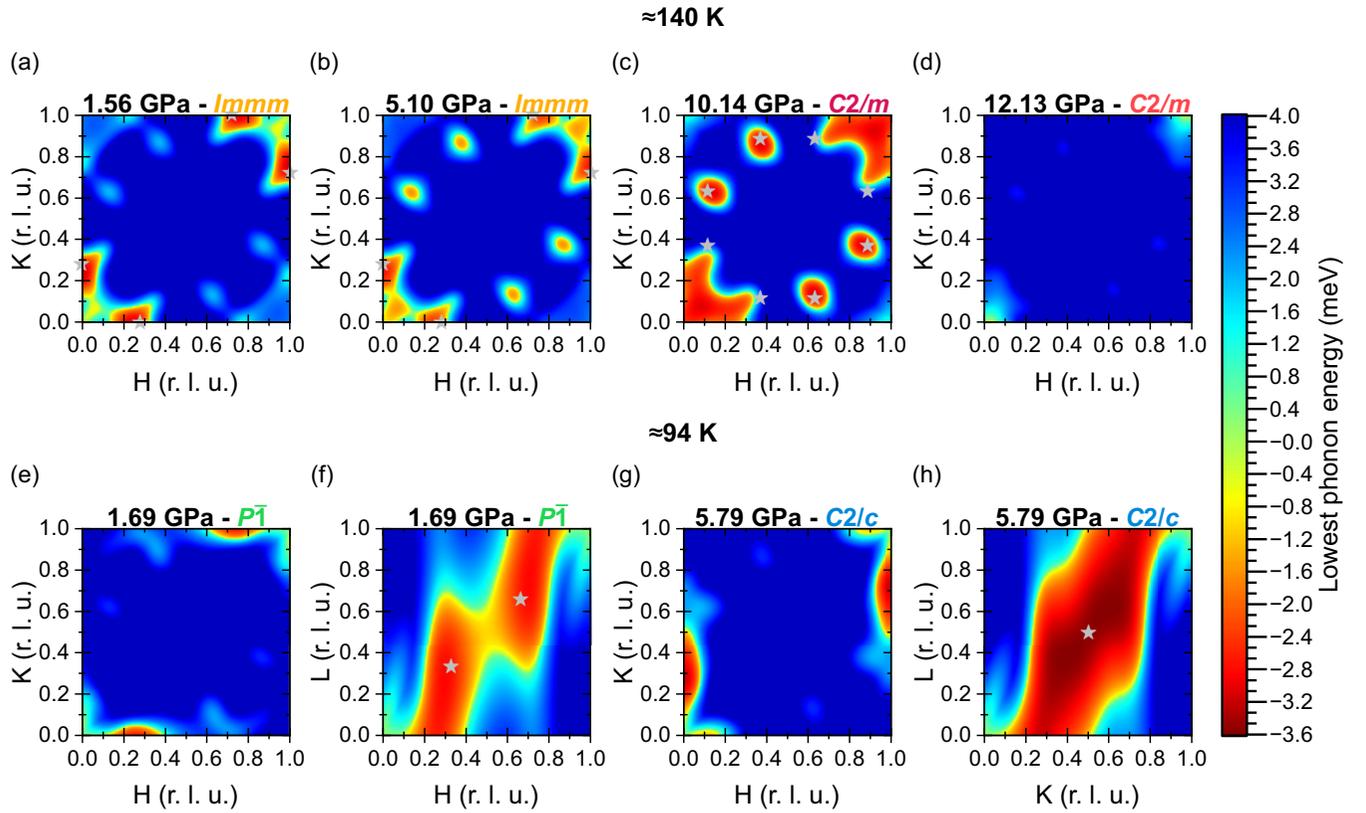

FIG. 5. Calculation of the phonon instabilities. ($HK0$) map of the lowest phonon frequency calculated by DFPT using the experimental crystal structures at 140 K and (a) 1.56, (b) 5.1, (c) 10.14, and (d) 12.13 GPa. The experimental structure for (d) was measured at 145 K. Map of the lowest phonon frequency calculated by DFPT using the experimental crystal structures at 94 K and 1.69 GPa in the (e) ($HK0$) and (f) ($H0L$) planes and at 5.79 GPa in the (g) ($HK0$) and (h) ($0KL$) planes. The gray stars indicate the experimentally obtained wave vector of the CDW observed at that pressure-temperature point.

## V. SUMMARY AND OUTLOOK

In summary, we investigated the pressure dependence of the crystal structure and CDWs of superconducting BaNi$_2$As$_2$ and revealed the formation of original structural polymorphs and CDWs. At high pressure, a monoclinic phase exhibiting planar Ni zigzags forms and is stable against CDW instabilities. A detailed phase diagram of BaNi$_2$As$_2$ was determined and revealed a highly unusual pressure dependence of the incommensurate and commensurate CDW phases of this compound. First-principles calculations fueled by the experimental crystal structures showed a series of lattice instabilities in very good agreement with the experimentally observed ones. The stable monoclinic high-pressure phase shows a strongly reduced interlayer As-As distance, bearing striking similarities to previously encountered collapsed tetragonal phases in pnictides, highlighting the importance of the hybridization between As and Ni orbitals in controlling the electronic phases of these compounds. This calls for additional investigations, in particular regarding the impact of the reported structural phase transitions on the superconducting transition temperature of BaNi$_2$As$_2$.

*Note added.* Recently, we became aware of another high-pressure study of BaNi$_2$As$_2$ [53]. In agreement with our findings, the structural analysis carried out on powder samples in that study reported a collapsed phase at high pressures, but in the absence of full structural refinements across the $p$-$T$ phase diagram, it did not identify the monoclinic distortion reported in the present work.

Data measured at ID15B (ESRF) are available from the European Synchrotron Radiation Facility [54], and the data measured at P02.2 (PETRA III, DESY) are available from the Karlsruhe Institute of Technology [55].

**ACKNOWLEDGMENTS**

We acknowledge DESY (Hamburg, Germany), a member of the Helmholtz Association HGF, for the provision of experimental facilities. Parts of this research were carried out at PETRA III using beamline P02.2. Beam time was allocated for Proposal No. I-20200263. We acknowledge the European Synchrotron Radiation Facility (ESRF) for provision of synchrotron radiation facilities under Proposal No. HC-4234, and we would like to thank D. Comboni and T. Poreba for assistance and support in using beamline ID15B. We acknowledge the funding from the Deutsche Forschungsgemeinschaft (DFG; German Research Foundation) Project ID 422213477-TRR 288 (Project No. B03) and support from the state of Baden-Württemberg through bwHPC. S.-M.S. acknowledges funding from the Deutsche Forschungsgemeinschaft [Project No. 441231589].

S.-M.S. and T.L. prepared the samples for the XRD measurements. T.L., S.-M.S., A.-A.H., G.G., and K.G. carried out the XRD experiments. A.-A.H., T.L., S.-M.S, and M.M. analyzed the XRD data. A.-A.H. solved (and refined) the crystal structures. A.-A.H. and T.L. grew the single crystals. R.H. performed first-principal calculations. S.-M.S., A.-A.H., and M.L.T. conceived and supervised the project. M.L.T., T.L., and A.-A.H. wrote the manuscript with input from all the coauthors.

[1] G. Lander, E. Fisher, and S. Bader, The solid-state properties of uranium a historical perspective and review, Adv. Phys. **43**, 1 (1994).

[2] B. Keimer, S. A. Kivelson, M. R. Norman, S. Uchida, and J. Zaanen, From quantum matter to high-temperature superconductivity in copper oxides, Nature (London) **518**, 179 (2015).

[3] K. Rossnagel, On the origin of charge-density waves in select layered transition-metal dichalcogenides, J. Phys.: Condens. Matter **23**, 213001 (2011).

[4] T. Neupert, M. M. Denner, J.-X. Yin, R. Thomale, and M. Z. Hasan, Charge order and superconductivity in kagome materials, Nat. Phys. **18**, 137 (2022).

[5] F. Ronning, N. Kurita, E. D. Bauer, B. L. Scott, T. Park, T. Klimczuk, R. Movshovich, and J. D. Thompson, The first order phase transition and superconductivity in $BaNi_2As_2$ single crystals, J. Phys.: Condens. Matter **20**, 342203 (2008).

[6] Y. Yao, R. Willa, T. Lacmann, S.-M. Souliou, M. Frachet, K. Willa, M. Merz, F. Weber, C. Meingast, R. Heid, A.-A. Haghighirad, J. Schmalian, and M. Le Tacon, An electronic nematic liquid in $BaNi_2As_2$, Nat. Commun. **13**, 4535 (2022).

[7] C. Meingast, A. Shukla, L. Wang, R. Heid, F. Hardy, M. Frachet, K. Willa, T. Lacmann, M. Le Tacon, M. Merz, A.-A. Haghighirad, and T. Wolf, Charge density wave transitions, soft phonon, and possible electronic nematicity in $BaNi_2(As_{1-x}P_x)_2$, Phys. Rev. B **106**, 144507 (2022).

[8] A. S. Sefat, M. A. McGuire, R. Jin, B. C. Sales, D. Mandrus, F. Ronning, E. D. Bauer, and Y. Mozharivskyj, Structure and anisotropic properties of $BaFe_{2-x}Ni_xAs_2$ ($x = 0, 1$, and $2$) single crystals, Phys. Rev. B **79**, 094508 (2009).

[9] S. Lee, G. de la Peña, S. X.-L. Sun, M. Mitrano, Y. Fang, H. Jang, J.-S. Lee, C. Eckberg, D. Campbell, J. Collini, J. Paglione, F. M. F. de Groot, and P. Abbamonte, Unconventional charge density wave order in the pnictide superconductor $Ba(Ni_{1-x}Co_x)_2As_2$, Phys. Rev. Lett. **122**, 147601 (2019).

[10] M. Merz, L. Wang, T. Wolf, P. Nagel, C. Meingast, and S. Schuppler, Rotational symmetry breaking at the incommensurate charge-density-wave transition in $Ba(Ni, Co)_2(As, P)_2$: Possible nematic phase induced by charge/orbital fluctuations, Phys. Rev. B **104**, 184509 (2021).

[11] M. Frachet, P. Wiecki, T. Lacmann, S. M. Souliou, K. Willa, C. Meingast, M. Merz, A.-A. Haghighirad, M. Le Tacon, and A. E. Böhmer, Elastoresistivity in the incommensurate charge density wave phase of $BaNi_2(As_{1-x}P_x)_2$, npj Quantum Mater. **7**, 115 (2022).

[12] C. Eckberg, D. J. Campbell, T. Metz, J. Collini, H. Hodovanets, T. Drye, P. Zavalij, M. H. Christensen, R. M. Fernandes, S. Lee, P. Abbamonte, J. W. Lynn, and J. Paglione, Sixfold enhancement of superconductivity in a tunable electronic nematic system, Nat. Phys. **16**, 346 (2020).

[13] S. M. Souliou, T. Lacmann, R. Heid, C. Meingast, M. Frachet, L. Paolasini, A.-A. Haghighirad, M. Merz, A. Bosak, and M. Le Tacon, Soft-phonon and charge-density-wave formation in nematic $BaNi_2As_2$, Phys. Rev. Lett. **129**, 247602 (2022).

[14] K. Kudo, M. Takasuga, Y. Okamoto, Z. Hiroi, and M. Nohara, Giant phonon softening and enhancement of superconductivity by phosphorus doping of $BaNi_2As_2$, Phys. Rev. Lett. **109**, 097002 (2012).

[15] T. Noda, K. Kudo, M. Takasuga, M. Nohara, T. Sugimoto, D. Ootsuki, M. Kobayashi, K. Horiba, K. Ono, H. Kumigashira, A. Fujimori, N. L. Saini, and T. Mizokawa, Orbital-dependent band renormalization in $BaNi_2(As_{1-x}P_x)_2$ ($x = 0.00$ and $0.092$), J. Phys. Soc. Jpn. **86**, 064708 (2017).

[16] C. Eckberg, L. Wang, H. Hodovanets, H. Kim, D. J. Campbell, P. Zavalij, P. Piccoli, and J. Paglione, Evolution of structure and superconductivity in $Ba(Ni_{1-x}Co_x)_2As_2$, Phys. Rev. B **97**, 224505 (2018).

[17] S. Lee, J. Collini, S. X.-L. Sun, M. Mitrano, X. Guo, C. Eckberg, J. Paglione, E. Fradkin, and P. Abbamonte, Multiple charge density waves and superconductivity nucleation at antiphase domain walls in the nematic pnictide $Ba_{1-x}Sr_xNi_2As_2$, Phys. Rev. Lett. **127**, 027602 (2021).

[18] L. Sang, Z. Li, G. Yang, Z. Yue, J. Liu, C. Cai, T. Wu, S. Dou, Y. Ma, and X. Wang, Pressure effects on iron-based superconductor families: Superconductivity, flux pinning and vortex dynamics, Mater. Today Phys. **19**, 100414 (2021).

[19] T. Park, H. Lee, E. D. Bauer, J. D. Thompson, and F. Ronning, Pressure dependence of $BaNi_2As_2$, J. Phys.: Conf. Ser. **200**, 012155 (2010).

[20] P. G. Naumov, K. Filsinger, O. I. Barkalov, G. H. Fecher, S. A. Medvedev, and C. Felser, Pressure-induced transition to the collapsed tetragonal phase in $BaCr_2As_2$, Phys. Rev. B **95**, 144106 (2017).

[21] M. Tomić, R. Valentí, and H. O. Jeschke, Uniaxial versus hydrostatic pressure-induced phase transitions in $CaFe_2As_2$ and $BaFe_2As_2$, Phys. Rev. B **85**, 094105 (2012).

[22] K. Syassen, Ruby under pressure, High Press. Res. **28**, 75 (2008).

[23] C. Prescher and V. B. Prakapenka, Dioptas: A program for reduction of two-dimensional x-ray diffraction data and data exploration, High Press. Res. **35**, 223 (2015).

[24] Rigaku Oxford Diffraction, CRYSALISPRO software system, version 1.171.42.57a, Rigaku Corporation, Wroclaw, Poland, 2022.

[25] See Supplemental Material at http://link.aps.org/supplemental/10.1103/PhysRevB.108.224115 for information on the samples, five isobars, additional calculations of the phonon linewidth, and joint density of states and more details on the observed experimental structures.

[26] G. M. Sheldrick, A short history of *SHELX*, Acta Cryst. A **64**, 112 (2008).

[27] G. M. Sheldrick, Crystal structure refinement with *SHELXL*, Acta Cryst. C **71**, 3 (2015).

[28] V. Petříček, M. Dušek, and L. Palatinus, Crystallographic computing system JANA2006: General features, Z. Kristallogr. **229**, 345 (2014).

[29] K. Momma and F. Izumi, *VESTA*: A three-dimensional visualization system for electronic and structural analysis, J. Appl. Cryst. **41**, 653 (2008).

[30] S. G. Louie, K.-M. Ho, and M. L. Cohen, Self-consistent mixed-basis approach to the electronic structure of solids, Phys. Rev. B **19**, 1774 (1979).

[31] B. Meyer, C. Elsässer, and M. Fähnle, Fortran90 program for mixed-basis pseudopotential calculations for crystals (unpublished).

[32] D. R. Hamann, M. Schlüter, and C. Chiang, Norm-conserving pseudopotentials, Phys. Rev. Lett. **43**, 1494 (1979).

[33] G. B. Bachelet, D. R. Hamann, and M. Schlüter, Pseudopotentials that work: From H to Pu, Phys. Rev. B **26**, 4199 (1982).

[34] D. Vanderbilt, Optimally smooth norm-conserving pseudopotentials, Phys. Rev. B **32**, 8412 (1985).

[35] J. P. Perdew, K. Burke, and M. Ernzerhof, Generalized gradient approximation made simple, Phys. Rev. Lett. **77**, 3865 (1996).

[36] R. Heid and K.-P. Bohnen, Linear response in a density-functional mixed-basis approach, Phys. Rev. B **60**, R3709(R) (1999).

[37] A. Kreyssig, M. A. Green, Y. Lee, G. D. Samolyuk, P. Zajdel, J. W. Lynn, S. L. Bud'ko, M. S. Torikachvili, N. Ni, S. Nandi, J. B. Leão, S. J. Poulton, D. N. Argyriou, B. N. Harmon, R. J. McQueeney, P. C. Canfield, and A. I. Goldman, Pressure-induced volume-collapsed tetragonal phase of $CaFe_2As_2$ as seen via neutron scattering, Phys. Rev. B **78**, 184517 (2008).

[38] S. Jia, A. J. Williams, P. W. Stephens, and R. J. Cava, Lattice collapse and the magnetic phase diagram of $Sr_{1-x}Ca_xCo_2P_2$, Phys. Rev. B **80**, 165107 (2009).

[39] B.-H. Lei, Y. Guo, Y. Xie, P. Dai, M. Yi, and D. J. Singh, Complex structure due to as bonding and interplay with electronic structure in superconducting $BaNi_2As_2$, Phys. Rev. B **105**, 144505 (2022).

[40] Z. P. Yin, S. Lebègue, M. J. Han, B. P. Neal, S. Y. Savrasov, and W. E. Pickett, Electron-hole symmetry and magnetic coupling in antiferromagnetic LaFeAsO, Phys. Rev. Lett. **101**, 047001 (2008).

[41] T. Yildirim, Strong coupling of the Fe-spin state and the As-As hybridization in iron-pnictide superconductors from first-principle calculations, Phys. Rev. Lett. **102**, 037003 (2009).

[42] D. A. Zocco, J. J. Hamlin, K. Grube, J.-H. Chu, H.-H. Kuo, I. R. Fisher, and M. B. Maple, Pressure dependence of the charge-density-wave and superconducting states in $GdTe_3$, $TbTe_3$, and $DyTe_3$, Phys. Rev. B **91**, 205114 (2015).

[43] A. Sacchetti, C. L. Condron, S. N. Gvasaliya, F. Pfuner, M. Lavagnini, M. Baldini, M. F. Toney, M. Merlini, M. Hanfland, J. Mesot, J.-H. Chu, I. R. Fisher, P. Postorino, and L. Degiorgi, Pressure-induced quenching of the charge-density-wave state in rare-earth tritellurides observed by x-ray diffraction, Phys. Rev. B **79**, 201101(R) (2009).

[44] M. Leroux, I. Errea, M. Le Tacon, S.-M. Souliou, G. Garbarino, L. Cario, A. Bosak, F. Mauri, M. Calandra, and P. Rodière, Strong anharmonicity induces quantum melting of charge density wave in $2H$-$NbSe_2$ under pressure, Phys. Rev. B **92**, 140303(R) (2015).

[45] S. Lee, T. B. Park, J. Kim, S.-G. Jung, W. K. Seong, N. Hur, Y. Luo, D. Y. Kim, and T. Park, Tuning the charge density wave quantum critical point and the appearance of superconductivity in $TiSe_2$, Phys. Rev. Res. **3**, 033097 (2021).

[46] S. Raymond, J. Bouchet, G. H. Lander, M. Le Tacon, G. Garbarino, M. Hoesch, J.-P. Rueff, M. Krisch, J. C. Lashley, R. K. Schulze, and R. C. Albers, Understanding the complex phase diagram of uranium: The role of electron-phonon coupling, Phys. Rev. Lett. **107**, 136401 (2011).

[47] F. H. Yu, D. H. Ma, W. Z. Zhuo, S. Q. Liu, X. K. Wen, B. Lei, J. J. Ying, and X. H. Chen, Unusual competition of superconductivity and charge-density-wave state in a compressed topological kagome metal, Nat. Commun. **12**, 3645 (2021).

[48] D. Song, Y. Zhou, M. Zhang, X. He, and X. Li, Structural and transport properties of 1T-$VSe_2$ single crystal under high pressures, Front. Mater. **8**, 710849 (2021).

[49] B. Woo, S. Seo, E. Park, J. H. Kim, D. Jang, T. Park, H. Lee, F. Ronning, J. D. Thompson, V. A. Sidorov, and Y. S. Kwon, Effects of pressure on the ferromagnetic state of the charge density wave compound $SmNiC_2$, Phys. Rev. B **87**, 125121 (2013).

[50] A. Subedi and D. J. Singh, Density functional study of $BaNi_2As_2$: Electronic structure, phonons, and electron-phonon superconductivity, Phys. Rev. B **78**, 132511 (2008).

[51] F. Weber, S. Rosenkranz, J.-P. Castellan, R. Osborn, R. Hott, R. Heid, K.-P. Bohnen, T. Egami, A. H. Said, and D. Reznik, Extended phonon collapse and the origin of the charge-density wave in $2H$-$NbSe_2$, Phys. Rev. Lett. **107**, 107403 (2011).

[52] F. Weber, S. Rosenkranz, J.-P. Castellan, R. Osborn, G. Karapetrov, R. Hott, R. Heid, K.-P. Bohnen, and A. Alatas, Electron-phonon coupling and the soft phonon mode in $TiSe_2$, Phys. Rev. Lett. **107**, 266401 (2011).

[53] J. Collini, D. J. Campbell, D. Sneed, P. Saraf, C. Eckberg, J. Jeffries, N. Butch, and J. Paglione, Charge order evolution of superconducting $BaNi_2As_2$ under high pressure, Phys. Rev. B **108**, 205103 (2023).

[54] T. Lacmann, S. M. Souliou, G. Gabarino, and M. Le Tacon, High pressure study of charge density wave ordering in $BaNi_2As_2$ (version 1) [Data set], 2023 data, European Synchrotron Radiation Facility, https://doi.org/10.15151/ESRF-DC-1305040552.

[55] T. Lacmann, S. M. Souliou, A.-A. Haghighirad, K. Glazyrin, and M. Le Tacon, High pressure study of charge density wave ordering in $BaNi_2As_2$ [Data set], 2023 data, Karlsruhe Institute of Technology, https://doi.org/10.35097/1656.